\documentclass[twoside]{article}
\usepackage{epsf,fleqn,espcrc2}
\begin{document}
\title{Scale Determination Using the Static Potential with Two Dynamical 
Quark Flavors }
\author{Sonali Tamhankar\thanks{presented by S.\ Tamhankar} and Steven Gottlieb \\ Indiana University, 
Bloomington, IN 47405, USA}
\begin{abstract}
We study the static potential using gauge configurations that include
the effects of two flavors of dynamical Kogut-Susskind quarks.
The configurations, generated by the MILC collaboration, and used to
study the spectrum and heavy-light decay constants,
cover a range $5.3 \le 6/g^2 \le 5.6$.  There are
at least four quark masses for each coupling studied.  Determination of
$r_0$ from the potential can be used to set a scale.  This alternative
scale is useful to study systematic errors on the spectrum and decay
constants.
\end{abstract}
\maketitle
\section{INTRODUCTION}

\renewcommand{\topfraction}{1}
\renewcommand{\bottomfraction}{1}
\renewcommand{\textfraction}{0.001}
To get an idea of systematic errors introduced in determining the physical 
scale in lattice QCD simulations using $m_\rho$, 
it is important to use an alternative quantity to set the physical scale. 
One quantity, $r_0$, proposed by Sommer
\cite{sommer} has recently been studied in several papers 
\cite{acc,heller,allton}. 
Here, we examine the static potential for a range of couplings
5.3 $\leq 6/g^2 \leq$ 5.6 with two flavors of dynamical staggered quarks.
We compute the Sommer scale for each dynamical quark mass and extrapolate to 
the chiral limit at each coupling. 
Finally, using the scale determined by $r_0$, we take the continuum limit
and replot the Edinburgh curve, finding very little difference as compared
with setting  the scale from $m_\rho$.

The configurations were generated by the MILC collaboration \cite{milc} 
using the staggered quark action with two dynamical flavors and the 
Wilson gauge action. 
They were stored every 10 units of molecular dynamics time. 
These configurations have been used elsewhere for light hadron 
spectrum calculations \cite{conf}.  
For each coupling there are at least four quark masses.
(For details, see Ref.~\cite{details}.)
\section{STATIC POTENTIAL}
\begin{figure}[thb]
\vspace{-0.5cm}
\epsfxsize=7.0cm
\epsfysize=6.0cm
\epsfbox{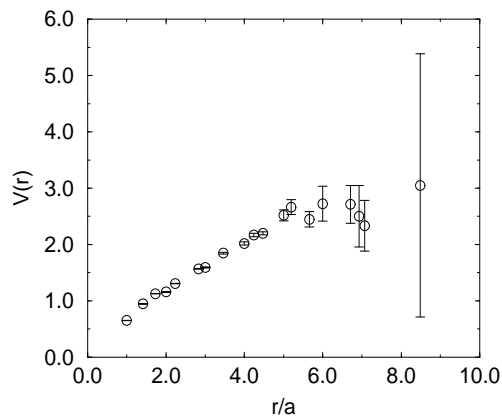}
\vspace{-1.0cm}
\caption{Hint of string breaking: flattening of the last few points.                     $\beta$ = 5.3, $am_q = 0.075$.}
\label{str_brk}
\end{figure}
To calculate the static potential,
Wilson loops $W(R,T)$ were measured for 5 and 10 smearing steps 
using the APE smearing method as explained in \cite{ape}. 
The effective potentials were then calculated using
\begin{equation}
V_T(R) = \ln \left[\frac{W(R,T)}{W(R,T+1)}\right]
\end{equation}
After blocking the results into 30 time unit blocks,
errors are obtained using single elimination jackknife.
\begin{figure}[t]
\vspace{1.5cm}
\epsfxsize=8.0cm
\epsfysize=14.0cm
\epsfbox{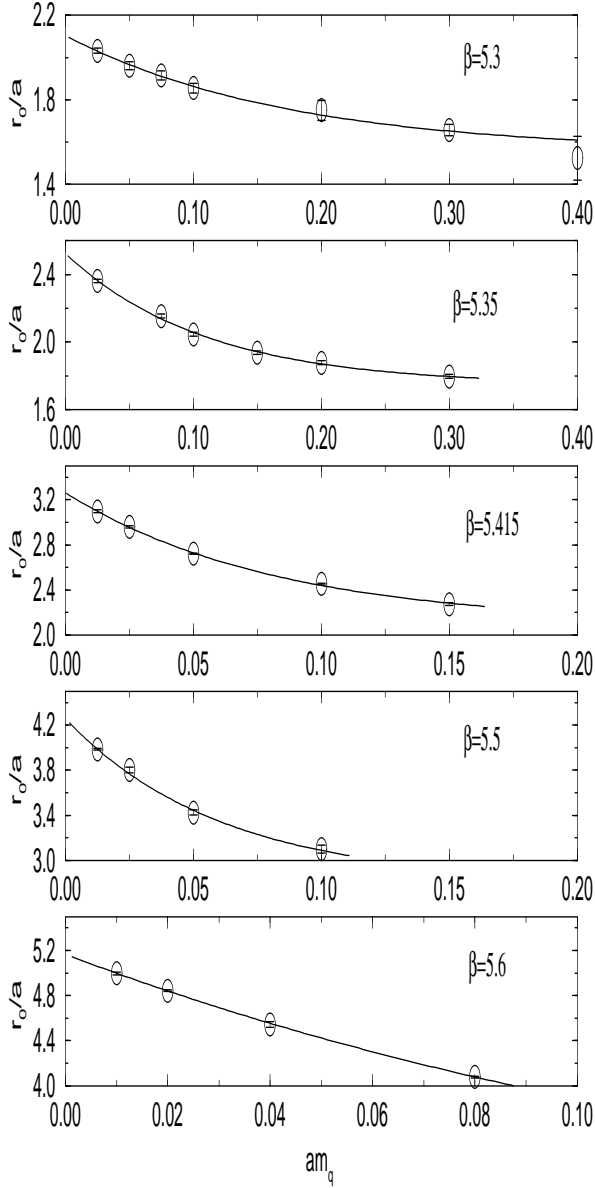}
\vspace{-0.5cm}
\caption{
The graphs show $r_0/a$ {\it vs}.\ $am_q$ for different $\beta$ values, along with an exponential fit defined in Eq.~(4).  }
\vspace{-1cm}
\label{six_graphs}
\end{figure}
$V(R)$ is expected to flatten out at large $R$ for dynamical quarks due 
to string breaking.  
At our strongest coupling, 5.3, and smallest quark masses we have
searched for evidence of string breaking. 
For the two lightest masses, we did not find a signal at large enough values 
of $R$, but we do find a hint of string breaking for $am_q$ = 0.075.  In Fig.~1,
the potential seems to flatten for $R$ greater than five to seven lattice 
spacings.  This corresponds to a distance of 1.3--1.8 fm. Ref. \cite{detar} reports finding signs of string breaking at 0.8--1.1 fm, however their $m_\pi/m_\rho$ is lower. This is consistent: the string is expected to break at smaller distances for lighter quarks.
\vspace{1cm}
\begin{figure}[htb]
\epsfxsize=7.0cm
\epsfysize=6.0cm
\epsfbox{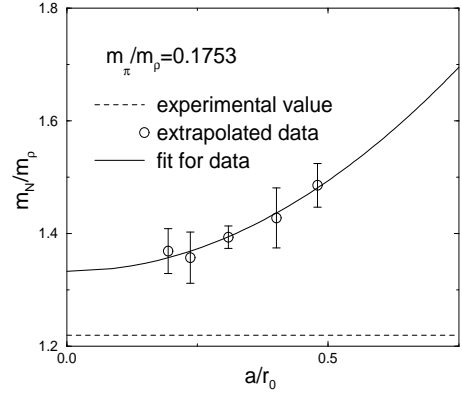}
\vspace{-1cm}
\caption{Continuum extrapolation for $m_\pi/m_\rho=0.1753$. Continuum extrapolation was made at various $m_\pi/m_\rho$ values to plot the Edinburgh curve shown in Fig.~4}
\end{figure}
\vspace{-1cm}
\section{FIT PARAMETERS}
We used the standard ansatz for the potential,
\begin{equation}
V(R)=V_0+\sigma R - e/R -f(G_L-1/R) .
\end{equation}
This is a simple combination of a Coulomb term at short distances and a linear 
increase at large distances. The last term accounts for lattice artifacts. 
For more details on this ansatz and its limitations see Refs.~\cite{acc}
and \cite{ansatz}.

Using the fit parameters $\sigma$ and $e$,
the Sommer scale $r_0$ is calculated in the usual way,
\begin{equation}
r_0=\sqrt{\frac{c-e}{\sigma}}
\end{equation}
with c = 1.65.  This corresponds to a physical length of $r_0 \sim 0.5$ fm.

We fit the potential to the ansatz (2) for different ranges of $R$ and 
different values of $T$. 
The "good" fits yielded consistent values, except for 
$\beta$ = 5.35, $am_q$=0.05. 
We have excluded this point from our fits for the chiral extrapolation. 
The criteria for best fits were the confidence of the fit 
and the degrees of freedom. 
Additionally, the range of $R$ had to include the value 
of $r_0$, which was an important constraint for the coarse lattices. 
In Fig.~2,
we show the best estimates of $r_0$ along with the exponential fit 
explained below.

\section{SCALE DETERMINATION}

For dynamical quarks, a different value of  $r_0$ and $\sigma$  is obtained for 
each quark mass.  The consideration that the physical scale should approach the 
quenched result as the quark mass tends to infinity and the graphs of $r_0$ 
vs $am_q$  suggest an exponential fit
\begin{equation}
r_0=A+B\left(e^{-Cam_q}-1\right)
\end{equation}
with three parameters $A$, $B$ and $C$.

We obtained good fits to this form for all $\beta$  values except 5.415, 
where we excluded the heaviest two points from the fit to get an 
acceptable confidence level. 
These fits were used for the chiral extrapolation to determine $r_0$ for each $m_q$.
\vspace{0.3cm}
\begin{figure}[thb]
\vspace{-0.5cm}
\epsfxsize=7.0cm
\epsfysize=6.0cm
\epsfbox{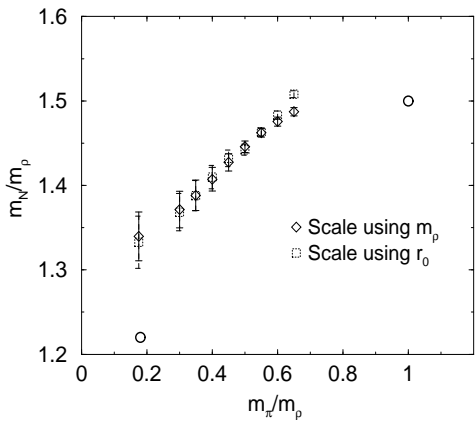}
\vspace{-1cm}
\caption{Continuum Edinburgh curves for dynamical fermions using the scale as
determined by $m_\rho$ and $r_0$.}
\label{trial}
\vspace{-1cm}
\end{figure}

\section{EDINBURGH PLOT}

A most surprising result of Lattice '98 was that the Edinburgh 
plot for dynamical
quarks was shifted up compared to the quenched simulations, leading to a larger 
deviation from the real world value in the continuum limit \cite{conf}. 
We were curious to know if the Edinburgh curve changed when setting the 
scale from $r_0$.
Fig.~3 shows $m_N/m_\rho$ as a function of $a/r_0$ for the 
real world $m_\pi/m_\rho$  
value.\footnote{These fits were good up to  $m_\pi/m_\rho$  = 0.45
but for higher ratios the $\beta$  = 5.415 point departed from the curve. There 
seems to be a problem with that coupling for higher masses that we do not 
completely understand yet.  Hence we have excluded that point from the fits 
for continuum extrapolation for  $m_\pi/m_\rho > $ 0.45. This exclusion yielded better fits without much shift in the continuum value.} 
The Edinburgh curve is replotted using the scale determined here in Fig. 4.

\section{CONCLUSIONS}

We have calculated the heavy quark potential for a range of couplings for 
dynamical staggered quarks. 
Some flattening of the static potential for large $R$ at
strong coupling is seen which hints at string breaking, 
but does not suffice to make a conclusive case. 
We determine $r_0$, extrapolate it to the chiral limit using 
an exponential form and examine the hadron mass data using the new scale. 
We do not find significant deviations from the 
previous continuum extrapolations using the $\rho$ mass to set the scale.

We would like to thank the rest of the MILC collaboration for the use of the 
lattices and U.M.~Heller for some fitting routines. We are grateful to the US 
Department of Energy which supported this work under grant FG 02-91ER-40661.
The computations were performed on the T3E at PSC, Origin-2000s at NCSA and 
an Origin-2000, a Paragon and the CANDYCANE Linux cluster at Indiana University.

\end{document}